\documentstyle[12pt]{article}
\newcommand{\be}{\begin{eqnarray}}
\newcommand{\ee}{\end{eqnarray}}
\newcommand{\no}{\nonumber}
\begin{document}
\begin{titlepage}
\vspace*{1.0cm}
\begin{center}
{\Large\bf Anyon in External Electromagnetic Field:\\ Hamiltonian
and Langrangian Formulations}
\vskip 1.0 cm
by
\vskip 1.0 cm
{\bf M. Chaichian}$^{1,2}$, {\bf R. Gonzalez
Felipe}$^{1,}$\renewcommand{\thefootnote}
{*}\footnote{On leave of absence from Grupo de Fisica Te\'{o}rica, ICIMAF,
Academia de Ciencias de Cuba, Calle E No. 309, Vedado, Habana 4, Cuba.},
{\bf D.Louis Martinez}$^{3,*,}$\renewcommand{\thefootnote}{\dagger}\footnote
{Present address: Department of Physics, Univ. of Manitoba, Winnipeg MB
R3T 2N2, Canada.}
\vskip 2.0 cm
\end{center}

$^1$High Energy Physics Laboratory, Physics Department, P.O. Box 9
(Siltavuorenpenger 20 C), SF-00014, University of Helsinki, Finland

$^2$Research Institute for High Energy Physics, P.O. Box 9 (Siltavuorenpenger
20 C), SF-00014, University of Helsinki, Finland

$^3$Research Institute for Theretical Physics, P.O. Box 9 (Siltavuorenpenger
20C), SF-00014, University of Helsinki, Finland
\vskip 2.5 cm
\noindent{\bf Abstract}

We propose a simple model for a free relativistic particle of fractional spin
in 2+1 dimensions which satisfies all the necessary conditions. The
canonical quantization of the system leads to the description of
one-particle states of the Poincar\'{e} group with arbitrary spin. Using the
Hamiltonian formulation with the set of constraints, we introduce the
electromagnetic interaction of a charged anyon and obtain the Lagrangian.
The Casimir operator of the extended algebra, which is the first-class
constraint, is obtained and gives the equation of motion of the anyon. In
particular, from the latter it follows that the gyromagnetic ratio for a
charged anyon is two due to the parallelness of spin and momentum of the
particle in 2+1 dimensions. The canonical quantization is also
considered in this case.
\end{titlepage}
\noindent{\bf 1. Introduction}
\vskip 0.5 cm
The existence of anyons or particles with arbitrary spin and statistics in
2+1 dimensions [1] has been attracting a great deal of  attention due
to the applications to different planar physical phenomena such as the
fractional quantum Hall effect and possibly, high-Tc superconductivity and
to the description of physical processes in the presence of cosmic strings.
Several field-theoretic models have been proposed, in which anyons appear
as topological solitons [2,3] or electrically charged vortices [4,5]. In
another
approach, point particles, described by scalar or spinor fields, are coupled
minimally to a $U(1)$ gauge field, the so-called statistical gauge field,
whose dynamics is governed by the Chern-Simons action [6]. However, none of
the above models gives a description for a free particle with arbitrary
spin.

In Refs. [7,8] the field equations for a free particle with fractional spin
were proposed and it was shown that their solutions realize the one-particle
states as the appropriate induced representation of the Poincar\'{e} group
in 2+1 dimensions. Besides, in [7] the corresponding classical action for the
fields was constructed by analogy with the action of the massive vector field.
Ref. [9] dealt with the same problem, but starting from the description of the
classical action for a relativistic particle with fixed mass and fixed
arbitrary spin. There the set of Hamiltonian constraints was found and two
different schemes of the canonical quantization of the model were considered.
However, no interaction of anyons with the electromagnetic field has been
considered in the above mentioned works.

In Ref. [10] based on heuristical arguments an equation for anyon in a
constant magnetic field was assumed for the description of relativistic
fractional quantum Hall effect. In Ref. [11], using the approach of
coupling fermions or bosons to a statistical Chern-Simons field, it was
shown that one-anyon states acquire an induced magnetic moment consistent
with a value of $g=2$ for the gyromagnetic ratio. In a recent paper [12]
the electromagnetic interaction of anyons has been considered on the
basis of classical analogy with the behaviour of spin in an electromagnetic
field and intuitive arguments.

In this letter we propose Hamiltonian and Lagrangian descriptions for both
a free and an interacting (with electromagnetic field) relativistic
particles with fractional spin. Our derivations are based on the imposed
constraints and the algebraic properties of the system, namely on the
invariants,
Casimir operators, of the corresponding extended algebra.
\vskip 1.0 cm
\noindent{\bf 2. Free relativistic particle with fractional spin}
\vskip 0.5 cm
We consider first a simple model of a relativistic particle in 2+1 dimensions,
described by the action
\be
I=\int L\ d\tau\ ,
\ee
where
\be
L=\frac{m(\dot{x}\dot{n})}{\sqrt{\dot{n}^2}}
\ee
is the Lagrangian, $\tau$ is an evolution parameter, $x^\mu$ are the
coordinates
of the particle, $\mu=0,1,2$, and the dot denotes the derivative with respect
to $\tau$. The auxiliary vector $n^\mu$ is a space-like unit vector
$(n^2=-1)$  which, as we shall
see below, serves to describe the spin degree of freedom of the particle.
We shall take the metric as $g_{\mu\nu}=diag(1,-1,-1)$. Finally, $m$ is a
parameter with dimension of mass. The constants $\hbar$ and $c$ are set to
unity.

Let us consider now the Hamiltonian formulation of the system in order
to show that our proposed Lagrangian (2) indeed describes a free relativistic
particle of mass $m$ with arbitrary spin. With this aim we introduce the
canonical momenta $p_\mu$ and $p_\mu^{(n)}$, conjugated to the generalized
coordinate $x_\mu$ and $n_\mu$ respectively, and which satisfy the canonical
Poisson brackets (PB):
\be
& &\{ x_\mu,p_\nu\}=-g_{\mu\nu}\ \  ,\ \ \ \{ n_\mu,p^{(n)}_\nu\}=-g_{\mu\nu}
\ ,\\
& &\{ x_\mu,x_\nu\}=\{ p_\mu,p_\nu\}=0\ \ ,\ \ \ \{ n_\mu, n_\nu\}=\{
p^{(n)}_\mu,p^{(n)}_\nu\}=0\ .\no
\ee
{}From the Lagrangian (2) we obtain the momenta:
$$p_\mu=\frac{\partial L}{\partial\dot{x}^\mu}=\frac{m\dot{n}_\mu}
{\sqrt{\dot{n}^2}}\ ,\hspace{1.5cm}\eqno{(4a)}$$
$$p^{(n)}_\mu=\frac{\partial L}{\partial\dot{n}^\mu}=\frac{m}{\sqrt{\dot{n}^2}}
\biggl(\dot{x}_\mu-\frac{(\dot{x}\dot{n})\dot{n}_\mu}{\dot{n}^2}\biggr)
\ .\eqno{(4b)}$$
Eqs. (4) together with the condition $n\dot{n}=0$ lead to the
primary constraints [13]
$$\Phi_1=p^2-m^2=0\ ,\eqno{(5a)}$$
$$\varphi_1=(pn)=0\ ,\hspace{0.3cm}\eqno{(5b)}$$
$$\varphi_2=(pp^{(n)})=0\ .\hspace{0.4cm}\eqno{(5c)}$$
The mass-shell condition (5a) is a first-class constraint (i.e. its PB's
with all the constraints of the system vanish), while (5b) and (5c) are
second-class constraints since $\{\varphi_1,\varphi_2\}=p^2=m^2\neq 0$.
It is also straightforward to prove that the canonical Hamiltonian
$H_{can}=\dot{x}p+\dot{n}p^{(n)}-L$ is equal to zero and consequently,
the total Hamiltonian of the system is a linear combination of the constraints
(5),
\addtocounter{equation}{+2}
\be
H=\Lambda_1\Phi_1+\lambda_1\varphi_1+\lambda_2\varphi_2\ .
\ee
where $\Lambda_1(\tau),\lambda_1(\tau),\lambda_2(\tau)$ are the Lagrange
multipliers ($\Lambda_i$ is associated with the first-class
and $\lambda_j$ with the second-class constraints, hereafter). Let us introduce
now the spin vector $S_\mu$ by analogy with the orbital angular momentum [9].
We shall define
\be
S_\mu=\varepsilon_{\mu\nu\lambda}n^\nu p^{(n)\lambda}\ ,
\ee
where $\varepsilon_{\mu\nu\lambda}$ is the totally antisymmetric tensor
$(\varepsilon_{012}=1)$.

{}From (7) it follows that
\be
\{ S_\mu,S_\nu\}=\varepsilon_{\mu\nu\lambda}S^\lambda\ .
\ee
The quantity $M_{\mu\nu}=x_\mu p_\nu-x_\nu p_\mu+n_\mu p^{(n)}_\nu-n_\nu
p^{(n)}_\mu$ is the conserved total angular momentum tensor. It is easy
to show that eqs. (5b) and (5c) imply that $\varepsilon_{\mu\nu\lambda}
S^\nu p^\lambda=0$, i.e. the spin vector $S^\mu$ is parallel to the momentum
$p_\mu$ on 2+1 dimensions. Thus, we can write
\be
S^\mu=-\frac{\alpha p^\mu}{\sqrt{p^2}}\ ,
\ee
where $\alpha$ is an arbitrary constant (in principle, it
could be a function of $\tau$, but from the equations of motion
(10) it follows that $\dot{\alpha}=0)$, which fixes the spin of the particle.
{}From the Hamiltonian (6) (or equivalently from the Lagrangian (2)) and the
definition (7) we obtain the following equations of motion:
\be
\dot{p}_\mu=0\ \ ,\ \ \ \dot{S}_\mu=0\ ,
\ee
i.e. the model describes a free relativistic particle of mass $m$ and with
any arbitrary spin due to the fact that the parameter $\alpha$ does not
appear explicitly in the Lagrangian. We can pass now to the Dirac quantization
[13] of the system described by the Hamiltonian (6) with the set of constraints
(5). The second-class constraints (5b) and (5c) define the so-called Dirac
brackets, which give the commutation rules:
\be
& & \ [x_\mu,x_\nu]=-i\varepsilon_{\mu\nu\lambda}\frac{S^\lambda}{p^2}\ ,\no\\
& & \ [x_\mu,p_\nu]=-ig_{\mu\nu}\ ,\\
& & \ [p_\mu,p_\nu]=0\ ,\no\\
& & \ [n_\mu,n_\nu]=0\  ,\no\\
& & \ [p^{(n)}_\mu,p^{(n)}_\nu]=0\ ,\no\\
& & \ [n_\nu,p^{(n)}_\nu]=-i\biggl( g_{\mu\nu}+\frac{1}{p^2}p_\mu p_\nu\
\biggr)\ .\no
\ee

The first-class constraint (5a) turns into the equation specifying the
physical quantum states of the system:
\be
(p^2-m^2)\psi=0\ .
\ee
Note also that the spin operator (7) commutes with the constraint (5a) and
therefore, it is a physical observable of the theory.

Now we shall consider the model in which the spin of the particle is fixed
(it enters explicitly as a parameter in the Lagrangian). With this aim, in
addition to constraints (5), we introduce a new first-class constraint
\be
\Phi_2=Sp+\alpha m=0\ ,
\ee
where $S_\mu$ is given by eq.(7)\renewcommand{\thefootnote}{*}{\footnote{
A different model with more constraints than (5) and (13) was proposed in [9],
leading to different Hamiltonian and Lagrangian. However, in that model
$\ddot{x}_\mu\neq 0$. In our proposed model $\ddot{x}_\mu=0$ and consequently,
$x_\mu$ describes the space-time coordinate of the free particle.}.
The first-class constraints (5a) and (13) are the invariants (Casimir
operators) of the Poincar\'{e} algebra in 2+1 dimensions [14,7].
The requirement (5a) is the mass-shell condition and eq. (13) specifies the
helicity, with $\alpha$ an (arbitrary) value of spin. Note also that the set
of constraints (5), (13) leads to the relation (9).

We take as the total Hamiltonian of the system the linear combination
of the primary constraints (5) and (13)
\be
H=\Lambda_1\Phi_1+\Lambda_2\Phi_2+\lambda_1\varphi_1+\lambda_2\varphi_2\ ,
\ee
(with $H_{can}=0$). Performing the inverse Legendre transformation, we find
the Lagrangian
\be
L=m\sqrt{\frac{(\varepsilon_{\mu\nu\lambda}\dot{x}^\mu n^\nu\dot{n}^\lambda-
\alpha\frac{\dot{n}^2}{m})^2+(\dot{x}\dot{n})^2}{\dot{n}^2}}\ .
\ee
By direct verification one can show that Lagrangian (15) leads to the set of
constraints (5) and (13) as primary constraints and that no secondary
constraints appear in the model. The corresponding canonical Hamiltonian
is equal to zero.

In the Dirac quantization scheme, the first-class constraints (5a) and (13)
define the equations for the physical quantum states of the system [7,9]
\be
(p^2-m^2)\psi=0\ \ ,\ \ \ (Sp+\alpha m)\psi=0\ ,
\ee
while the second-class constraints (5b), (5c) give the commutation rules (11).
Since under quantization no restrictions on the parameter $\alpha$ appear, the
physical states described by the wave functions (16), are the states of a
particle with mass $m$ and arbitrary spin $\alpha$. It is straightforward
to verify that the equations of motions (10) are also satisfied in this case.

Thus we have shown that Lagrangians (2) and (15) describe a free relativistic
particle with arbitrary spin, with the only difference that in the case of
eq. (2), the spin of the particle is not fixed in the theory, while Lagrangian
(15) depends explicitly on the value of spin $\alpha$.
\vskip 1.0 cm
\noindent{\bf 3. Anyon in external electromagnetic field}
\vskip 0.5 cm
Let us study now the problem of introducing the interaction of
electromagnetic field with a fractional spin particle. By analogy with
the noninteracting case [7], where the one-particle states are specified
by values assigned to the invariants (Casimir operators) of the Poincar\'{e}
algebra, we shall define the single-anyon states in a constant electromagnetic
field as the unitary representations of the relevant operator algebra in
2+1 dimensions. For a uniform constant electromagnetic field, the corresponding
extended algebra has the form:
\be
 \ [S_\mu,S_\nu]&=&i\varepsilon_{\mu\nu\lambda}S^\lambda\ ,\no\\
 \ [S_\mu,\pi_\nu]&=&i\varepsilon_{\mu\nu\lambda}\pi^\lambda\ ,\no\\
 \ [S_\mu,\tilde{F}_\nu]&=&i\varepsilon_{\mu\nu\lambda}\tilde{F}^\lambda\ ,\\
 \ [\pi_\mu,\pi_\nu]&=&ie\varepsilon_{\mu\nu\lambda}\tilde{F}^\lambda\ ,\no\\
 \ [\pi_\mu,\tilde{F}_\nu]&=&0\ ,\no\\
 \ [\tilde{F}_\mu,\tilde{F}_\nu]&=&0\ ,\no
\ee
where $\pi_\mu=p_\mu-eA_\mu,\ A_\mu=-\frac{1}{2}F_{\mu\nu}x^\nu$ is the vector
potential, $\tilde{F}_\lambda=\frac{1}{2}\varepsilon_{\lambda\mu\nu}F^{\mu\nu}$
and $F_{\mu\nu}$ is the uniform constant electromagnetic field tensor,
$F_{\mu\nu}=\partial_\mu A_\nu-\partial_\nu A_\mu$.

We can prove that the operator $\pi^2-2e(\tilde{F}S)$ is the Casimir of
the algebra (17). Therefore, in the classical theory we can take the quantity
$\pi^2-2e(\tilde{F}S)-m^2$ as the corresponding first-class constraint. We
will suppose this constraint to be valid even for a {\it nonuniform}
field\renewcommand{\thefootnote}{*}\footnote{This assumption leads to the
correct equations of motions in a nonuniform external electromagnetic field
(see eqs. (22)).} $\tilde{F}_\mu$ and replace the set of constraints (5) by the
following one:
$$\Phi_1=\pi^2-2e(\tilde{F}S)-m^2=0\ ,\eqno{(18a)}$$
$$\varphi_1=(\pi n)=0\ ,\hspace{1.2cm}\eqno{(18b)}$$
$$\varphi_2=(\pi p^{(n)})=0\ ,\hspace{1.0cm}\eqno{(18c)}$$
where $S_\mu$ is defined as in (7). The second-class constraints (18b),
(18c) imply that the spin vector $S_\mu$ is parallel to the momentum $\pi_\mu$
and we can write
\addtocounter{equation}{+1}
\be
S_\mu=-\alpha\frac{\pi_\mu}{\sqrt{\pi^2}}\ .
\ee

Now we define the total Hamiltonian of the system as follows
\be
H=\Lambda_1\Phi_1+\lambda_1\varphi_1+\lambda_2\varphi_2\ ,
\ee
where $H_{can}$ is taken to be equal zero; $\Phi_1,\varphi_1,\varphi_2$ are
given by eqs.(18) and $\Lambda_1,\lambda_1,\lambda_2$ are the Lagrange
multipliers. In this case, unlike the free one, we can not find the exact
analytic expression for the Lagrangian, since the equations for the Lagrange
multipliers $\Lambda_1$ and $\lambda_{1,2}$ become polynomials of high
degrees. However, keeping only the terms up to
linear approximation in the field
strength $\tilde{F}_\mu$, we can perform the inverse Legendre transformation
to obtain the following Lagrangian\renewcommand{\thefootnote}{**}\footnote{
In addition, this linear approximation in the field strength provides us
with the possibility to compare our results with the ones obtained
previously in the literature in the same approximation.}:
\be
L=\frac{m(\dot{x}\dot{n})}{\sqrt{\dot{n}^2}}\biggl( 1+\frac{e}{m^2}(\tilde{F}
S)\biggr) +eA\dot{x}\ ,
\ee
where $S_\mu$ is defined in (7) with $p^{(n)}_\mu$ given by (4b). The equations
of motion, which follow from (20) (or, equivalently (21)) read as
$$\dot{\pi}_\mu=\frac{e}{m}F_{\mu\nu}\pi^\nu+\frac{e}{m}S^\nu\partial_\mu
\tilde{F}_\nu\ ,\eqno{(22a)}$$
$$\dot{S}_\mu=\frac{e}{m}F_{\mu\nu}S^\nu\ ,\hspace{1.0cm}\eqno{(22b)}$$
which are the ones obtained in [12] by following a different approach. The
first
term in the r.h.s. of eq. (22a) is the Lorentz force, while the second one
corresponds to the coupling of the dipole moment to the gradient of the
field. Eq.(22b) is the Bargmann-Michel-Telegdi equation in 2+1 dimensions
for the precession of the spin in an external electromagnetic field [15].
{}From the latter equation it follows that the gyromagnetic ratio for anyons
is $g=2$ (see also [11], [12]), a fact which is a direct consequence of the
form of the Casimir operator of the algebra (17). Note also that we
can come to the same conclusion from the system of constraints (18). Indeed,
if one takes $\Phi_1$ in the form $\Phi_1=\pi^2-2e\eta(\tilde{F}S)-m^2=0$,
with $\eta$ a constant, it turns out that $\Phi_1$ is a first-class constraint
if and only if $\eta=1$. Thus, $g=2$ due to the parallelness [12] of spin and
momentum in 2+1 dimensions (guaranteed in our case by the second-class
constraints (18b), (18c)) and due to the fact that $\Phi_1$ should be a
first-class constraint.

Let us remark that the equations of motion are obtained in the form (22)
since on the mass-shell (i.e. when the equations of motion are taken into
account) $\dot{n}_\mu$ is parallel to $\dot{x}_\mu$.

It is interesting to notice that if we identify the time-like vector
$\dot{n}_\mu$
with $\dot{x}_\mu$ by setting $\dot{n}_\mu=-\dot{x}_\mu$, the Lagrangian (21)
coincides with the standard Lagrangian (in this case in any dimension), for
a spinless charged particle in an electromagnetic field
$$L=-m\sqrt{\dot{x}^2}+eA\dot{x}\ .$$

Let us consider now the quantization of the system, following the Dirac
method [13]. The second-class constraints (18b), (18c) can be used to obtain
up to terms linear in $\tilde{F}_\mu$ the following commutation rules:
\addtocounter{equation}{+1}
\be
& & \ [x_\mu,x_\nu]=-i\varepsilon_{\mu\nu\lambda}\frac{S^\lambda}{\pi^2}
\biggl( 1-e\frac{(\tilde{F}S)}{\pi^2}\biggr)\ ,\no\\
& & \ [x_\mu,\pi_\nu]=-ig_{\mu\nu}\biggl(1-e\frac{(\tilde{F}S)}{\pi^2}\biggr)
+ie\frac{\tilde{F}_\mu S_\nu}{\pi^2}\ ,\no\\
& & \ [\pi_\mu,\pi_\nu]=ie\varepsilon_{\mu\nu\lambda}\tilde{F}^\lambda=
ieF_{\mu\nu}\ ,\\
& & \ [n_\mu,n_\nu]=0\ ,\no\\
& & \ [p^{(n)}_\mu,p^{(n)}_\nu]=0\ ,\no\\
& & \ [ n_\mu,p^{(n)}_\nu]=-ig_{\mu\nu}-i\frac{\pi_\mu\pi_\nu}{\pi^2}
\biggl(1-e\frac{(\tilde{F}S)}{\pi^2}\biggr)\ .\no
\ee
The first-class constraint (18a) is imposed as an operator on the physical
quantum states giving the equation:
\be
(\pi^2-2e(\tilde{F}S)-m^2)\psi=0
\ee
This equation was heuristically assumed in [10] for an anyon in a constant
magnetic field in the context of relativistic fractional quantum Hall effect
and recently it was obtained in [12] by introducing the minimal coupling with
the
electromagnetic field in the sympletic structure. As it was shown in [12],
the nonrelativistic limit of eq. (24) gives for the magnetic moment of the
anyon $\mu=-\frac{e\alpha}{m}$ and we see again that $g=2$.

It might be interesting to mention here that, in the linear approximation
in the field $\tilde{F}_\mu$, the Lagrangian (21) can be obtained from the
free Lagrangian (2) by the following substitution:
\be
& &\dot{x}_\mu\rightarrow\dot{x}_\mu+\frac{e}{m}\frac{(\dot{x}\dot{n})}{\sqrt{
\dot{n}^2}}A_\mu\ ,\no\\
& & \dot{n}_\mu\rightarrow\dot{n}_\mu+\frac{e}{m}\sqrt{\dot{n}^2}A_\mu-
\frac{e}{m}\frac{(\dot{x}\dot{n})}{\sqrt{\dot{n}^2}}F_{\mu\nu}n^\nu\ .
\ee
Finally, some remarks are in order concerning the possible extension of the
model described by Lagrangian (15) to the case with electromagnetic
interaction.
The introduction of a new constraint $\Phi_2=S\pi+\alpha m=0$ together with
eqs. (18) will lead to a contradictory system of
constraints\renewcommand{\thefootnote}{*)}\footnote{The new
constraint $\Phi_2=S\pi+\alpha m=0$ can not
be first-class, since not all its PB's with the other constraints (18),
namely with (18b) and (18c), vanish. Considering it as second-class, would
form together with (18b) and (18c) a total odd number (three) of second-class
constraints which is not possible, unless one adds another second-class
constraint.} and besides, as we have seen before, the algebra (17) has
only one Casimir operator, namely, $\Phi_1$ given in (18a). In order to
obtain a consistent theory, where the spin parameter $\alpha$ appears
explicitly
in the Lagrangian, we are forced to modify the set of constraints. This
question
is under study.

We can also treat the electromagnetic field as a dynamical variable by
adding to the Lagrangian (21) the usual term $-\frac{1}{4}F_{\mu\nu}F^{\mu\nu}$
and construct the quantum electrodynamics for anyons. Another interesting
problem is to formulate the quantum version of the theory in terms of path
integrals.

\vskip 1.0 cm
{\bf Acknowledgements}

This work started in collaboration with Alejandro Cabo and Hugo Perez
Rojas. We are grateful to them for many important discussions and comments
at an early stage of the work. It is our pleasure to thank I. Komarov
and C. Montonen for
several clarifying discussions. One of us (R.G.F.) would like to thank the
ICSC-World Laboratory for financial support. D.L.M. would like to thank
the Centre for International Mobility (CIMO), Finland, for providing
with financial support.
\vskip 9.0cm

\pagebreak

\end{document}